\documentstyle[12pt,aasms,flushrt]{article}

\def\gta{\gtrsim}
\def\lta{\lesssim}

\def\be{\begin{equation}}
\def\ee{\end{equation}}

\def\omit#1{}

\begin{document}
\title{\bf Migrating Planets}

\medskip

\author{N. Murray$^1$, B. Hansen$^1$, M. Holman$^1$, and S. Tremaine$^{1,2}$}

\medskip

\affil{$^1$Canadian Institute for Theoretical Astrophysics, \\
 University of Toronto, Toronto, Ontario M5S 3H8, Canada}

\medskip

\affil{$^2$Canadian Institute for Advanced Research,\\
Program in Cosmology and Gravity}

\medskip

\begin{abstract}
A planet orbiting in a disk of planetesimals can experience an
instability in which it migrates to smaller orbital
radii. Resonant interactions between the planet and
planetesimals remove angular momentum from the planetesimals,
increasing their eccentricities.  Subsequently, the planetesimals
either collide with or are ejected by the planet, reducing the
semimajor axis of the planet.  If the surface density of
planetesimals exceeds a critical value, corresponding to $\sim0.03$
solar masses of gas inside the orbit of Jupiter, the planet will
migrate inward a large distance. This instability may explain the
presence of Jupiter-mass objects in small orbits around nearby
stars.
\end{abstract}

\bigskip

\noindent
In the standard theory of solar system formation, solid material orbiting in a
gaseous disk accumulates to form small rocky or icy bodies called
planetesimals (\cite{lissauer}). Protoplanets then form by accretion of
planetesimals. If a protoplanet accretes roughly ten Earth masses
($10M_\oplus$), it can then capture a gas envelope from the protoplanetary
disk and become a gas giant like Jupiter. We now know of a number of
Jupiter-mass objects orbiting solar-type stars well inside the radius where
rocky material can condense (Table 1; ref. {\it 2}-{\it 8}). 
For example, a planet
orbits $\tau$ Bootis at a distance of $0.0462$ astronomical units (AU), where
the equilibrium temperature of 1550K is higher than the condensation
temperature of the most refractory minerals.  Although some of these
detections are controversial (\cite{gray}), we feel it is likely that most or
all are real. It is difficult to understand how to form such planets in place.

Although it is difficult to form planets at such small radii, once in place
they can survive (\cite{lin}). Thus it is natural to ask whether giant planets
can form at orbital radii of a few AU and then migrate inward. One proposed
migration mechanism involves the generation by the planet of density
waves in the gaseous protoplanetary disk, which cause the planet to spiral
inward (\cite{lin}, \cite{GT}). The movement of the
planet might be halted by short-range tidal or magnetic effects from the
central star (\cite{lin}); however, short-range stopping mechanisms cannot
easily explain the objects in Table 1 with  semimajor axes
$a_p\gta 0.2\hbox{ AU}$. 

Another migration scenario involves interactions between two or more
Jupiter-mass planets (\cite{RF}, \cite{WM}). If two such
planets form with a small enough separation their orbits are unstable, resulting in
either a collision or an ejection.  The typical outcome is a system with a
massive planet in an eccentric orbit at 1--2 AU, similar to 70 Vir or HD
114762. In a tiny fraction of cases the periapsis of one of the planets will
come close enough to the star that tidal evolution circularizes the orbit,
resulting in a massive planet on a circular orbit inside 0.1 AU. As pointed
out in (\cite{WM}) this low yield is difficult to reconcile with the observed
frequency of such systems, roughly a few percent of nearby stars. A second
difficulty is the low eccentricity and relatively large $a_p$ of the
planet orbiting $\rho$ Corona Borealis, which would require 
an additional damping mechanism.

In this paper we discuss a migration mechanism in which the planet exchanges
energy and angular momentum with the residual planetesimal disk through
resonant gravitational interactions, gravitational scattering, and physical
collisions. Long-distance migration is triggered (the ``migration
instability'') if the surface density $\Sigma$ of planetesimals exceeds a
critical value $\Sigma_c$ derived below. The migration halts when $\Sigma$ drops
below $\Sigma_c$, or when a significant fraction of scattered planetesimals
strike the star. Since planetesimals cannot survive inside a few stellar radii
($R_*$), the migration is certain to halt. If the mass of the planet is less
than about three times the mass of Jupiter ($M_J$), the eccentricity of the
planet's orbit is reduced by the migration; if not, the eccentricity may
increase.

The orbit of a planetesimal is specified by its semimajor axis $a$,
eccentricity $e$, and inclination $i$; the corresponding elements for the
planet are $a_p$, $e_p$, and $i_p$. In our analytic work, but not in
out numerical work, we assume that the mutual inclination
between the orbits of planetesimals and planet is $\lta0.1$ radians. It will
also be convenient to use the specific energy $E=-GM_*/(2a)$ and angular
momentum $L=[GM_* a(1-e^2)]^{1/2}$, where $M_*$ is the mass of the central
star; we shall not distinguish between the total angular momentum $L$ and its
normal component $L\cos i$. We assume that the ratio $\mu=M_p/M_*$ of planet
mass to stellar mass is much less than unity ($1/\mu=M_\odot/M_J\simeq1047$
for Jupiter), as is the ratio of planetesimal mass to planet mass.

Planet formation remains poorly understood, but we flank this obstacle
by focusing on processes that occur after formation is nearly
complete. We shall assume for simplicity that only a single massive planet
forms. The newly formed planet is probably surrounded by an annular
gap in the planetesimal disk, whose radial extent is $\sim 2.5$ times the
Hill radius, $h=\mu^{1/3}a_p$ (\cite{lissauer}). However, a planet's
gravitational reach can exceed its Hill radius grasp by a larger
factor, as we now explain.

Consider a planetesimal with orbital period 
$T=2\pi(a^3/GM_*)^{1/2}$ and $e\ll1$,
perturbed by an exterior planet with period $T_p>T$, $a_p=a+\Delta a$ and
$e_p\sim e$.  If the planetesimal is in a mean-motion resonance, so that
$(k+q)T\simeq kT_p$, where $k$ and $q$ are positive integers, the planet can
force chaotic perturbations of the planetesimal orbit. The semimajor axis $a$
is nearly constant, since the planetesimal is in resonance, but the
eccentricity $e$ undergoes a random walk, gradually diffusing to larger
values. Eventually the planetesimal crosses the planet's orbit, after a time
(\cite{HM})
\be 
T_R\sim{(2\Delta a)^{2q-1}\over\mu}{(e_me)^{2-\kappa}\over e_p^{2(q-\kappa)}},
\label{eq:tr}
\ee 
where $\kappa\in[0,q]$ and $q\gta2$. This time is given in units in which the
gravitational constant $G$, $M_*+M_p$, and $a_p$ are equal to unity, so
$T_p=2\pi$. The minimum $e$ at which the planetesimal can enter the
Hill sphere of the planet is $e_m\simeq(a_p-h)/a-1$ (\cite{small_q}). For a
Jupiter-mass planet at $a_p=5.2$ AU, and a planetesimal in the 7:4 resonance
with initial eccentricity $e_0=0.05$, we find $T_R\approx0.5\times10^5$ yr
(\cite{other_res}).

The orbital phase of a planet-crossing planetesimal at times of conjunction
with the planet is effectively random because of resonance overlap. Thus close
encounters between the two objects are common. The first close encounter
generally removes the planetesimal from the resonance, but leaves it in a
planet-crossing orbit.  A planet-crossing planetesimal undergoes a random walk
in both $e$ and $a$ (or in $L$ and $E$), but the Jacobi constant $J=E-L$ is
roughly constant (\cite{jacoby}).  Scaling arguments suggest that the time to
random walk out of the system should generally be between $T_p/\mu$ and
$T_p/\mu^2$, corresponding to a range $10^4$--$10^7$ yr for Jupiter. This
range is consistent with the median lifetime of Jupiter-family comets,
$5\times10^5$ yr (\cite{levdun}), and with the lifetimes found in our
numerical simulations. For example, if we start a planetesimal in the 7:4
resonance it evolves at roughly constant $E$ or $a$ 
 while $e$ random walks from $0.05$ to $\sim0.3$ over
$\sim3.5\times10^5$ yr (Fig. 1).  At this point it
suffers a close encounter with Jupiter. It then follows curves of constant $J$
until ejected after about $4\times10^5$ yr. 

Planetesimals may also suffer fates other than ejection:

\begin{enumerate}

\item
Collision with the star: If the planetesimal random walks to
sufficiently small $L$ (large $e$), it will be absorbed by the star.  This channel
can become important once the planet has migrated to $a_p\lta5$--$10R_*$.
Planetesimals on resonant orbits with $a\lta(a_p-h)/2$ also tend to
collide with the star, because these become highly eccentric before
they become planet-crossing. 

\item
Collision with the planet: A planetesimal may collide
with the planet before being ejected. The probability
is $1-\exp(-P)$, where  
\be 
\label{collide}
P\simeq {2M_*r_p\over M_pa_p}
\simeq 2\times10^{-3}(\mu^2\rho)^{-1/3}\left(5.2\hbox{ AU}\over a_p\right),
\ee 
where $\rho$ is the mean density of the planet. This result shows that
about $20\%$ of orbit-crossing planetesimals will 
strike Jupiter but most will be ejected (\cite{velocities}).

\item
Long-term capture into mean-motion resonances: Planetesimals on
planet-crossing orbits can be captured temporarily into resonances; this is
the inverse of the process described by equation (\ref{eq:tr}), which now
gives the typical residence time in the resonance. The median residence time
is generally short compared to the age of the solar system. However, a few
planetesimals can be trapped for very long times near stable islands
(\cite{karney}).  Such trapping has been seen in numerical integrations of
planet-crossing orbits (\cite{duncan}).

\end{enumerate}

Planetesimals start with negative energy and are ejected with positive energy;
thus the ejection process must remove orbital energy from the planet, which
moves closer to the star (\cite{star_note}).  This phenomenon is well-known in
the context of our solar system: Fern\'andez and Ip (\cite{FI}) suggested that
the ejection of planetesimals caused the orbit of Jupiter to shrink by
0.1--0.2 AU. This shrinkage can naturally explain the depletion of the
outer asteroid belt (\cite{HM},
\cite{LR}).  If the surface density of planetesimals is above the 
critical value $\Sigma_c$, this process is unstable.  To calculate $\Sigma_c$,
we let the semimajor axis of the planet shrink by $\Delta
a_p$, reducing its energy by
\be 
\Delta E_p\sim\left|{GM_*M_p\Delta a_p\over 2a_p^2}\right|.
\ee 
Planetesimals newly captured into chaotic resonances will be
removed from the system, either by ejection or by consumption by the planet or
the star. The mass of planetesimals affected is
\be 
\Delta M\simeq 2\pi p_c\alpha a_p\Sigma(\alpha a_p)\left 
|\alpha \Delta a_p^{(0)}\right |,
\ee 
where $\alpha<1$ is a measure of the average $a/a_p$ of
the affected planetesimals (\cite{alpha}), and $p_c\approx 1$ is the
capture probability.  In disposing of a
mass $\Delta M$ of planetesimals the planet loses energy 
\be 
\Delta E_\alpha\sim f(a_p){GM_*\Delta M\over2\alpha a_p},
\ee 
where $f$ is the fraction of the original planetesimal energy that is taken
from the planet's orbit; we expect that $f\approx1$ for $a_p\gg R_*$
and becomes negative as $a_p\to R_*$, where the planet perturbs most
planetesimals into the star.

The migration process is unstable if $|\Delta  E_\alpha|>|\Delta E_p|$,
which requires that the density exceed the critical density $\Sigma_c$: 
\be 
\label{critical}
\Sigma(\alpha a_p) \gta\Sigma_c={M_p\over2\pi a_p^2\alpha f(a_p)p_c};
\label{eq:instab}
\ee 
in other words if the mass in the planetesimal disk interior to the planet is
of order $M_p$, the planet can migrate nearly to the surface of the star.

The migration halts when either (i) the local surface density
of planetesimals falls below the critical value, or (ii) a significant
fraction of the planetesimals plunge into the star.  The local surface density
is sure to fall to zero near the star. Solid bodies cannot
condense at radii $\lta 7R_*$, and existing planetesimals whose orbits might
evolve to smaller radii cannot long survive at distances $\lta 2R_*$. The
minimum semimajor axis achievable by the migration instability is thus a few
to ten stellar radii or $\sim0.03$--$0.1$ AU.

We have simulated the evolution of a
Jupiter-mass body in a planetesimal disk using the \"Opik approximation
(\cite{opik}); see Fig. 2. The simulations
assume that the surface density in planetesimals varies as
$\Sigma(r)=\Sigma_\odot(1\hbox{ AU}/r)^{1.5}$, where $\Sigma_\odot$, the
surface density at 1 AU, is a free parameter. The total mass in the
planetesimal disk within radius $r$ is then 
\be 
\label{total_mass}
M_D\simeq 1.4\times10^{-3} M_\odot (\Sigma_\odot/10^3\hbox{ g
cm}^{-2})(r/\hbox{ AU})^{0.5},
\ee 
and assuming solar metal abundance ($Z=0.02$) the
disk mass in gas interior to Jupiter's orbit is
$0.16M_\odot(\Sigma_\odot/10^3\hbox{ g cm}^{-2})$. 

The energy transfer from planet to planetesimal is approximated as a
succession of close encounters until the planetesimal is ejected, strikes the
planet or strikes the star. Summing over many encounters, we derive the
average efficiency of energy transfer from the planet to planetesimals in a
given resonance. These efficiency factors are then used to calculate the
evolution of the planet in a disk of a given initial mass profile.  For the
case where the planet interacts with the 2:1, 3:2, 2:3 and 1:2 resonances, as
well as a broad resonance zone in the immediate vicinity of the planet, the
onset of the migration instability is at $\Sigma_\odot\approx200\hbox{ g
cm}^{-2}$ (Fig. 2). Equation (\ref{critical}), evaluated with $a_p=5.2$ AU,
$\alpha=0.7$, $p_c=1$, and $f=1$  gives
$\Sigma_c\approx70\hbox{ g cm}^{-2}$, corresponding to
$\Sigma_\odot\approx160\hbox{ g cm}^{-2}$, in good agreement with our
numerical estimate.  A much higher density $\Sigma_\odot\simeq 8000\hbox{ g
cm}^{-2}$ is required for the instability to persist until the planet reaches
0.03 AU. 

In some cases a gas disk with the surface density
required for the migration instability may be nearly gravitationally
unstable. The criterion for local gravitational instability of a thin disk is
$Q=c\Omega/(\pi G\Sigma)<1$ where $c$ is the sound speed and $\Omega$ is the
angular speed (\cite{unstable}). Replacing $\Sigma$ by $\Sigma_c$ from
equation (\ref{critical}) we find
\be
Q=1.2\left(r\over1\hbox{ AU}\right)^{1/2}\left(T\over300\hbox{
K}\right)^{1/2}\left(Z\over0.02\right)\left(M_*\over M_\odot\right)^{1/2}
\left(M_J\over M_p\right),
\ee
where $T$ is the disk temperature. 

Ejecting a planetesimal removes energy $\Delta E$ and angular momentum
$\Delta L$ from the planet.  If $|\Delta E/\Delta L|>1$, then  $e_p$
decreases, while $e_p$ increases if $|\Delta E/\Delta L|<1$. For a
planet with $M_p\lta3M_J$, $|\Delta E/\Delta L|>1$ 
because the planetesimal must approach within a few Hill radii
of the planet to be ejected, and the Hill sphere does not extend
into the region where $|\Delta E/\Delta L|<1$ when $E\approx0$. In this case
the ejection of small planetesimals tends to reduce $e_p$ and $a_p$.
However the planet is immersed in a bath of planetesimals, many
of which have substantial masses. Interactions between the planet and
these objects will tend to produce equipartition between the energy in
radial motion of the planet and the planetesimals. If most of the
planetesimal mass is in objects of mass near $m_0$, the expected
equilibrium eccentricity of the planet is
$e_{eq}\approx(m_0/M_p)^{1/2}$. This ranges from 0.06--0.006 for $m_0$
between an Earth mass and a lunar mass.

Seven objects in Table 1 have $a_p<1$ AU, and probably did
not form in their present orbits.  We suggest that at least four of
these seven, 51 Pegasus, 55 Cancri, $\rho$ Corona Borealis, and $\tau$
Bootis, migrated to their present semimajor axes through the process
described here.  The planet around $\upsilon$ Andromeda is puzzling
because of its rather high eccentricity, $e_p=0.15\pm0.04$. Such an
eccentricity could arise from an encounter with an object of a
few $M_\oplus$, or the measured eccentricity could be in error.  The
last two short-period objects on the list, 70 Virginis with $M=6.6M_J$
and HD 114762 with $M=10\pm1M_J$, are sufficiently massive that
planetesimals scattered from their Hill sphere can be ejected with
$|\Delta L|>|\Delta E|$. If so, their large eccentricities could arise
from the migration instability.  Their relatively large $a_p$
might reflect the difficulty of satisfying the instability criterion
(\ref{eq:instab}) at small radii when $M_p$ is large.

The planet orbiting 47 Ursae Majoris could have formed at its
present location; however, we think it more likely that it migrated inward a
few AU and halted because $\Sigma(r)$ dropped below $\Sigma_c$.  The
companion to 16 Cygni B is also likely to 
have migrated inward; its large $e_p$ could be due to interactions
with the stellar companion to 16 Cygni B (\cite{HTT}).

Equation (\ref{total_mass}) shows  that a
planetesimal disk mass of $\sim 6\times10^{-4}M_\odot$ within the planetary
orbit is required to initiate the migration instability for a planet with
Jupiter's mass and radius, while a much larger mass of $\sim
2\times10^{-2}M_\odot$ is required to move the planet to 0.03 AU. Infrared
observations of solar-mass T Tauri stars suggest that disks at the upper end
of this mass range are
rare, but they do occur (\cite{BSCG}). These observations measure the total
mass in particulate matter at $r<100$ AU, with results in the range
$10^{-5}-10^{-2}M_*$ (\cite{gravitation}). The mass in planetesimals would be
similar if the efficiency of conversion of dust particles to planetesimals is
high, and even higher if the observed disks already hide most of their solid
material in planetesimal-sized objects.

There is weak observational evidence for a correlation between
the presence of short-period planets and high stellar 
metallicity (\cite{metal}). We note two possible explanations for this
correlation: (i) for a fixed mass of gas, enhancing the metallicity
increases the surface density of planetesimals, increasing the
chances for planet migration; (ii) if migration is halted by planetesimals
hitting the star, then the planetesimals can pollute the surface
layers of the star. 
 
With several planets migration becomes more complex; for example, in the solar
system Uranus and Neptune migrate outwards rather than inwards, because
Jupiter acts like an inner absorbing boundary similar to a nearby stellar
surface (\cite{FI}). In such a system the migration instability is triggered
at a lower surface density, since the outer planets effectively push the inner
planet toward the star.  The migration of a massive planet will reduce the
surface density of planetesimals substantially; as a result it is unlikely
that two massive planets can migrate to short-period orbits. It is
also unlikely that any other planets in the migration path can 
survive; thus a short-period Jupiter-mass planet should have no sister
planets with orbital radii less than a few AU.

\vfil\eject
\begin{planotable}{ccccc}
\tablecaption{Properties of planets}
\tablehead{
\colhead{Star}		& \colhead{Period (days)}	&
\colhead{$a_p$ (AU)} & \colhead{$M_p\sin i_p$ ($M_J$)}   & \colhead{$\ \ \ \ \ \ \ e_p$}
}
\startdata
$\tau$ Bootis (\cite{bmwhs})		&$3.3128\pm0.0002$	&$0.0462$	&$3.87$ &$0.018\pm0.016$\nl
$51$ Pegasi (\cite{mayor})		&$4.229\pm0.001$	&$0.05$		&$0.47$	&$0.0$\nl
$\upsilon$ Andromedae (\cite{bmwhs})	&$4.611\pm0.005$	&$0.057$	&$0.68$	&$0.15\pm0.04$\nl
$55$ Cancri	(\cite{bmwhs})		&$14.648\pm0.0009$	&$0.11$		&$0.84$	&$0.051\pm0.013$\nl
$\rho$ Corona Borealis (\cite{noyes})	&$39.645\pm0.88$	&$0.23$		&$1.1$	&$0.028\pm0.04$\nl
HD $114762$ (\cite{latham})		&$84.05\pm0.08$		&$0.3$		&$10\pm1$&$0.25\pm0.06$\nl
$70$ Virginis (\cite{marcy})		&$116.6$		&$0.43$		&$6.6$	&$0.4$\nl
$16$ Cygni B (\cite{cochran})		&$804$			&$1.7$		&$1.5$	&$0.67$\nl
$47$ Ursae Majoris (\cite{butler})	&$1090$			&$2.11$		&$2.39$	&$0.03\pm0.006$

\end{planotable}

\vfil\eject

\vfil\eject

Fig. 1. The evolution of the $i(t)$, $a(t)$ and $e(t)$
of a planetesimal having a mass $10^{-2}$ that of Jupiter.  The planetesimal
was initially placed in the 7:4 resonance with $e=0.05$ and $i=0.05$. The
perturbing planet had mass $\mu=10^{-3}$ and $e_p=0.05$. The planetesimal
suffers its first close encounter with the planet after $3.5\times10^5$ yr,
and is ejected after about $3.8\times10^5$ yr. 

Fig. 2.  The solid curves show the evolution of a 1 Jupiter-mass
planet in a disk with planetesimal surface density $\Sigma(r) =
\Sigma_\odot(1\hbox{ AU}/r)^{3/2}$.  We take initial elements as
follows; the plane has $a_p=5.2$ AU and $e_p=0.048$, while the
planetesimals have $a$ ranging from $0.03$ to $4$ AU, $e$ in the range
$(e_m,e_m+0.2)$, and $i$ between $0$ and $0.5$ radians.  The labelled
curves correspond to the following values of $\Sigma_\odot$ in units
of $\hbox{g cm}^{-2}$: A) 8000, B) 2000, C) 1200, D) 600, and E) 40
(the nominal value for the minimum solar nebula). The corresponding
gas surface densities in g cm$^{-2}$, assuming solar metallicity
$Z=0.02$, are A) $4\times10^5$, B) $1\times10^5$, C) $6\times10^4$, D)
$3\times10^4$, and E) $2\times10^3$.  There is a critical value of
$\Sigma_{\odot}$ at $\sim 200\hbox{ g cm}^{-2}$ (between D and E),
below which there is very little movement.  The solid circles indicate
the positions of the various extrasolar planets (Table~1); the J
indicates the location of Jupiter.  The dotted line (labelled $\rm
A^*$) corresponds to the case where we replaced the 2:1 resonance with
the 3:1 resonance, in which the planetesimals are not strictly
orbit-crossing unless one takes into account the finite Hill sphere
radius.  This indicates the approximate range of uncertainty in the
final position in this idealized model. The time for the migration to
occur is very model dependent; including more resonances will reduce
the migration time.


\begin{thebibliography}{}
\bibitem[{{\it 15}}]{small_q}
15. For resonances
with $q\le2$, $e_m$ is replaced by $e$, but this expression must be
used with caution when the resonances are strong, because the dynamics
are more complicated; see (\cite{HM}).
\bibitem[{{\it 16}}]{other_res}
16. Somewhat longer times apply in other low-order
mean-motion resonances. Similar results apply if the planet orbits
inside the planetesimal.
\bibitem[{{\it 17}}]{jacoby}
17. If the planetary eccentricity $e_p$ is non-zero, there is a random walk in
$J$, but this is much slower (by a factor of order
$e_p$) than the random walk in $E$ or $L$.
\bibitem[{{\it 22 }}]{star_note} 
22. By similar arguments, orbit-crossing planetesimals that collide with the
star add energy to the planet; however, consumption by the star is much less
common than ejection for planets at semimajor axes larger than a few stellar
radii.
\bibitem[{{\it 25}}]{alpha}
25. The average radius $\alpha=(a/a_p)=(7/4)^{-2/3}\simeq 0.7$ if
the inner 7:4 resonance is the only relevant one.
\bibitem[{{\it 30}}]{gravitation} 
30. A possible concern with the interpretation
of these observations is that the gas surface density may imply
gravitational instability if the disk extends out to 100 AU. See the
references in (27).
\end{thebibliography}
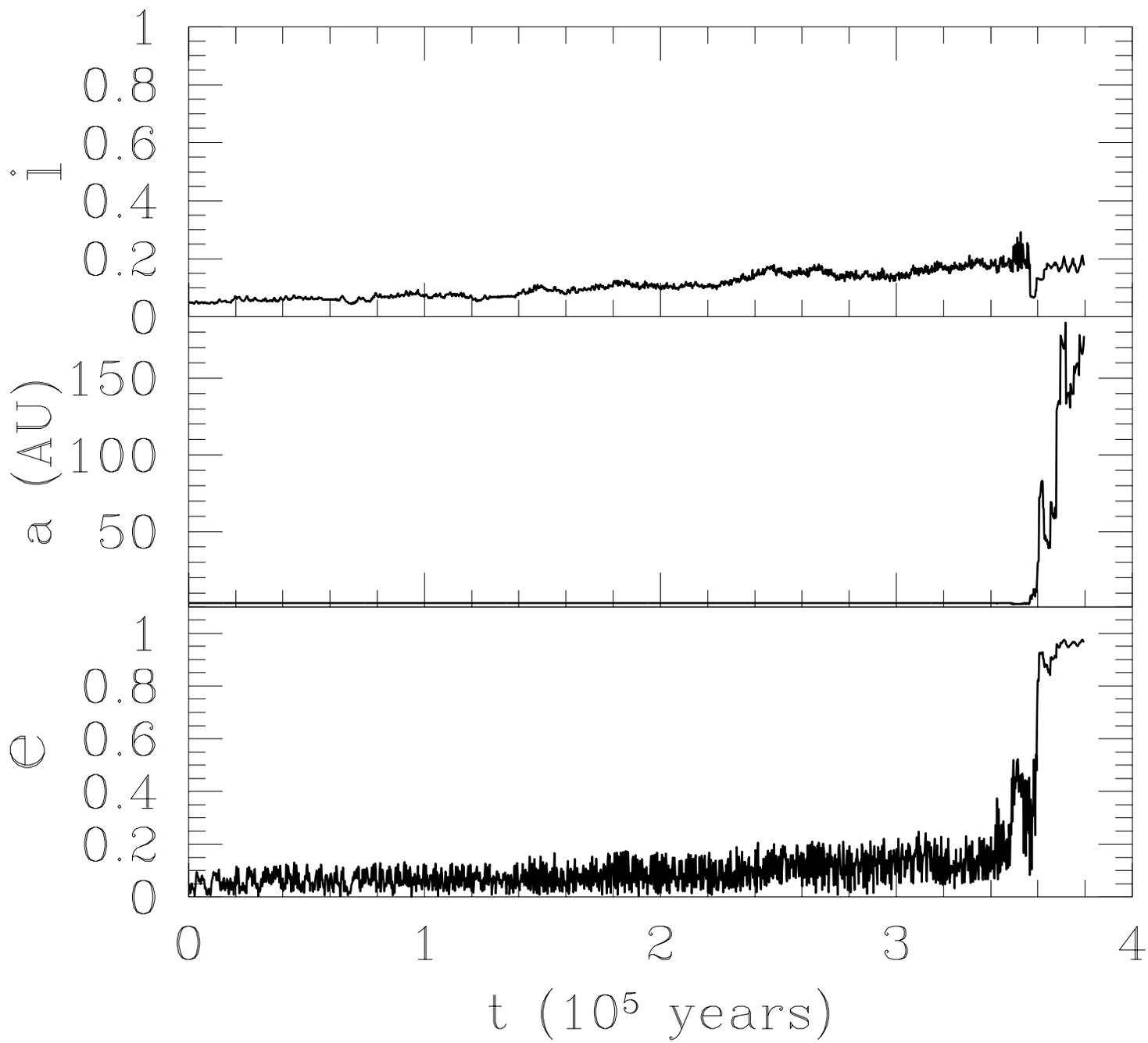
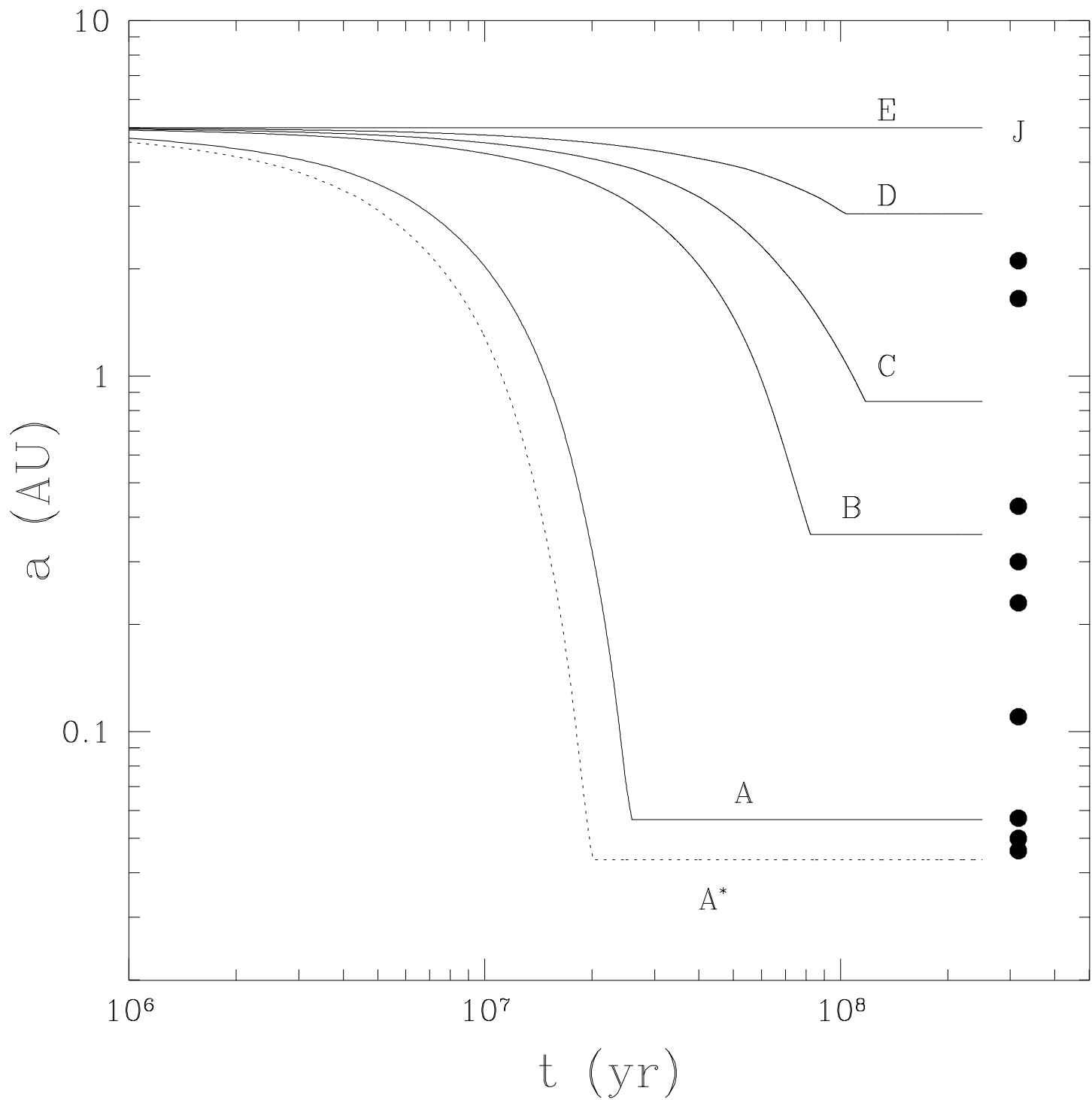
\end{document}